\documentclass[conference]{IEEEtran}
\IEEEoverridecommandlockouts

\newcommand{\conftitle}{17th International Workshop on Science Gateways (IWSG2025), 17-19 June 2025}

\usepackage{fancyhdr}
\usepackage{url}
\usepackage{url}
\usepackage{xcolor}
\usepackage{subcaption}
\usepackage{listings}
\usepackage{tikz}
\usetikzlibrary{shapes.geometric, arrows}

\fancypagestyle{pageStyle}{%
    \fancyhf{}
    \fancyhead[C]{\conftitle}
}

\newif\ifproofread

\newcommand{\changemarker}[1]{%
\ifproofread
\textcolor{blue}{#1}%
\else
#1%
\fi
}


  \usepackage{graphicx}

\usepackage{listings}%
\usepackage[most]{tcolorbox}%

\proofreadfalse

\begin{document}

\title{Interactive High-Performance Visualization\\ for Astronomy and Cosmology}
\renewcommand\IEEEkeywordsname{Keywords}

%\author{\IEEEauthorblockN{Eva Sciacca, Valentina Cesare,\\ Fabio Vitello, Ugo Becciani}
%\IEEEauthorblockA{INAF \\Astrophysical Observatory of Catania, \\Via Santa Sofia 78, Catania, Italy}
%\and
%\IEEEauthorblockN{Carmelita Carbone}
%\IEEEauthorblockA{INAF \\Institute of Space Astrophysics and Cosmic Physics, \\Via Alfonso Corti 12, Milano, Italy}
%\and
%\IEEEauthorblockN{Nicola Tuccari}
%\IEEEauthorblockA{University of Catania, \\Computer Science Department,\\ Viale Andrea Doria 6, Catania, Italy}
%\and
%\IEEEauthorblockN{Iacopo Colonnelli}
%\IEEEauthorblockA{University of Turin, \\Computer Science Department\\
%Via Pessinetto 12, Torino, Italy}}


\author{\IEEEauthorblockN{Eva Sciacca\IEEEauthorrefmark{1}\IEEEauthorrefmark{5}, Nicola Tuccari\IEEEauthorrefmark{2}\IEEEauthorrefmark{1}\IEEEauthorrefmark{5}, 
Umer Arshad\IEEEauthorrefmark{3}\IEEEauthorrefmark{5},
%Mattia D'Emidio\IEEEauthorrefmark{3},\\
%Gabriele Di Stefano\IEEEauthorrefmark{3},
Fabio Pitari\IEEEauthorrefmark{4},
Giuseppa Muscianisi\IEEEauthorrefmark{4}
, Emiliano Tramontana\IEEEauthorrefmark{2}
}
\IEEEauthorblockA{\IEEEauthorrefmark{1}INAF Astrophysical Observatory of Catania, Via Santa Sofia 78, Catania, Italy}
\IEEEauthorblockA{\IEEEauthorrefmark{2}University of Catania,  Department of Mathematics and Informatics, Viale Andrea Doria 6, Catania, Italy}
\IEEEauthorblockA{\IEEEauthorrefmark{3}University of L'Aquila, Department of Information Engineering, Computer Science and Mathematics, Via Vetoio, L'Aquila, Italy}
\IEEEauthorblockA{\IEEEauthorrefmark{4}Cineca, Via Magnanelli 6/3, 40033 Casalecchio di Reno, Bologna, Italy}
\IEEEauthorblockA{\IEEEauthorrefmark{5} All these authors contributed equally to the work}}



\maketitle
\thispagestyle{pageStyle}
\pagestyle{fancy}
\renewcommand{\headrulewidth}{0pt} % no line in header area

\begin{abstract}
The exponential growth of data in Astrophysics and Cosmology demands scalable computational tools and intuitive interfaces for analysis and visualization. In this work, we present an innovative integration of the VisIVO scientific visualization framework with the InterActive Computing (IAC) service at Cineca, enabling interactive, high-performance visual workflows directly within HPC environments. Through seamless integration into Jupyter-based science gateways, users can now access GPU-enabled compute nodes to perform complex 3D visualizations using VisIVO via custom Python wrappers and preconfigured interactive notebooks. We demonstrate how this infrastructure simplifies access to advanced HPC resources, enhances reproducibility, and accelerates exploratory workflows in astronomical research. Our approach has been validated through a set of representative use cases involving large-scale simulations from the GADGET code, highlighting the effectiveness of this system in visualizing the large-scale structure of the Universe. This work exemplifies how science gateways can bridge domain-specific tools and advanced infrastructures, fostering user-centric, scalable, and reproducible research environments.
\end{abstract}


% \fi
\IEEEpeerreviewmaketitle

% Code style
\definecolor{shellbg}{RGB}{45,9,34}
\definecolor{shellframe}{rgb}{.23,.23,.23}
\lstdefinestyle{bashscriptstyle}{
    language=bash,
    basicstyle=\scriptsize\color{white}\ttfamily\bfseries,
    breaklines=true,
}
\newtcblisting{bashscript}{
      top=0mm,
      bottom=0mm,
      left=0mm,
      right=0mm,
      boxrule=1pt,
      colback=shellbg,
      colframe=shellframe,
      title=bash,
      fonttitle=\bfseries\ttfamily\footnotesize\centering,
      listing only,
      listing options={style=bashscriptstyle},
}

% python style
\definecolor{pythonbg}{gray}{0.95}
\lstdefinestyle{pythonscriptstyle}{
    language=Python,
    backgroundcolor=\color{pythonbg},
    basicstyle=\ttfamily\footnotesize,
    keywordstyle=\color{blue}\bfseries,
    commentstyle=\color{green!50!black},
    stringstyle=\color{orange},
    breaklines=true,
    %frame=single,
    columns=flexible,
    captionpos=b,
    keepspaces=true,
}
\newtcblisting{pythonscript}{
      top=0mm,
      bottom=0mm,
      left=0mm,
      right=0mm,
      boxrule=1pt,
      boxsep=1pt,
      colback=pythonbg,
      colframe=shellframe,
      title=python,
      fonttitle=\bfseries\ttfamily\footnotesize\centering,
      listing only,
      listing options={style=pythonscriptstyle},
}

\section{Introduction}

Enormous data volumes, on the scale of petabytes, are produced through astrophysical observations or simulation codes running on high-performance supercomputers. These vast quantities of data create substantial obstacles to storage, retrieval, and analysis, which are crucial in facilitating scientific breakthroughs \cite{lan2021}.
%et: given the technological advances since 2009 a more recent paper would be better
%%
%%
Pre-exascale systems provide remarkable possibilities for scaling high-performance computing applications in Astrophysics and Cosmology (A\&C), necessitating both superior computational performance and interactive visualization of the results.

VisIVO\footnote{VisIVO, \url{https://visivo.readthedocs.io/}}, the Visualization Interface for the Virtual Observatory, is a set of tools designed for analyzing multi-dimensional data and uncovering previously unidentified connections within complex, multi-variate astrophysical datasets. It has been implemented in two ways: (1) through Science Gateways \cite{becciani2015science} for accessing Distributed Computing Infrastructures (DCIs) such as clusters, grids, and clouds, and (2) by leveraging containerization and virtualization technologies on the European Open Science Cloud (EOSC) infrastructure, integrated into interactive notebook applications \cite{sciacca2022scientific}.

In this study, we discuss the innovative incorporation of VisIVO into the InterActive Computing (IAC\footnote{IAC, \url{https://jupyter.g100.cineca.it/}}) service \cite{10.1007/978-3-030-82427-3_6} provided by the Cineca HPC centre.
IAC service allows end users to obtain access to HPC compute nodes via the web browser, replacing the traditional approach to HPC resources which is limited to a command line interface via \textit{ssh} and a queued batch system. Typical usage scenarios for IAC  are interactive analyses, visualizations, and steering of simulations running on scalable compute services that abstract large-scale computing resources, which can be used for running highly parallel simulation applications, but are also suitable for data analysis tasks, involving extreme-scale data sets.

The adopted strategy consists of the integration of VisIVO with IAC thus permitting its interactive execution on an HPC cluster (currently Galileo 100\footnote{Galileo 100 infrastructure: \url{https://www.hpc.cineca.it/systems/hardware/galileo100/}} at Cineca), with the goal of simplifying user engagement with the clusters and fostering reproducibility of the visualization workflows through the developed interactive notebooks.

\changemarker{In conclusion, we present the interactive notebooks developed through various VisIVO workflows, which aid in examining the Universe's large-scale structure. These studies utilize extensive cosmological N-body simulations generated with the GADGET code \cite{springel_05,viel_10}.}

\section{VisIVO Server}

VisIVO Server is a suite of software tools designed to create customized 3D visualizations from astrophysical data, supporting large-scale datasets without fixed limits on dimensionality.

Its main modules available via Command Line Interface (CLI) are as follows:

\begin{itemize}
    \item VisIVO Importer: converts user-supplied datasets into VisIVO Binary Tables (VBT), an efficient internal data representation. It supports various formats, including the general purpose data formats such as ASCII or CSV or tailored astronomical data formats such as FITS, HDF5, GADGET and more.
    \item VisIVO Filter: processes VBTs to apply data transformations, filtering, and other operations to prepare datasets for visualization.
    \item VisIVO Viewer: generates interactive 3D visualizations from VBTs, allowing users to explore and analyze their data effectively.
\end{itemize}

A typical usage of VisIVO Server involves at least three steps for data preparation, processing and visualization (see e.g. Figure \ref{fig:M1}). For sample VisIVO commands please see Section \ref{sec:workflows}.

\paragraph{Data Preparation}

Use VisIVO Importer to convert your dataset into a VBT. 

\paragraph{Data Processing (Optional)}

Apply one or more VisIVO Filter operations to perform data transformations or filtering on the VBT as needed.

\paragraph{Data Visualization}

Utilize one or more VisIVO Viewer renderings to create 3D visualizations from the (processed) VBT. 

This modular workflow allows for efficient handling and visualization of complex astrophysical datasets.

% Define block style
\tikzstyle{block} = [rectangle, draw, text centered, minimum height=1cm, minimum width=3cm]
\tikzstyle{data-block} = [cylinder, draw, shape border rotate=90, minimum height=0.5cm, minimum width=3cm]
\tikzstyle{user-block} = [ellipse, draw, text centered, minimum height=1cm, minimum width=3cm]

\tikzstyle{arrow} = [thick,->,>=stealth]

% VisIVO Server Architecture Diagram
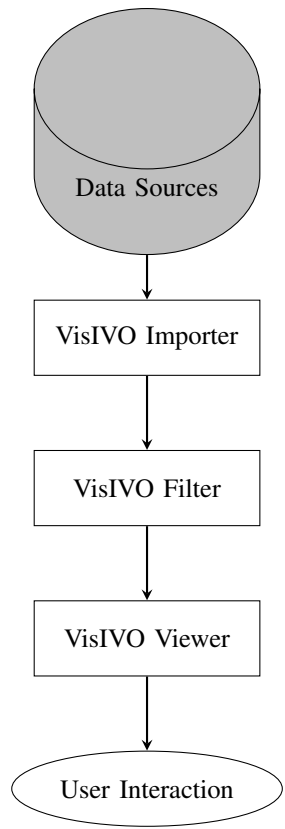
\begin{figure}
\begin{center}
\begin{tikzpicture}[node distance=2cm]
    \node (data) [data-block, fill=lightgray] {Data Sources};
    \node (importer) [block, below of=data] {VisIVO Importer};
    \node (filter) [block, below of=importer] {VisIVO Filter};
    \node (viewer) [block, below of=filter] {VisIVO Viewer};
    \node (user) [user-block, below of=viewer] {User Interaction};
    
    % Arrows
    \draw [arrow] (data) -- (importer);
    \draw [arrow] (importer) -- (filter);
    \draw [arrow] (filter) -- (viewer);
    \draw [arrow] (viewer) -- (user);
\end{tikzpicture}
\end{center}
\caption{VisIVO Server modular architecture and basic workflow involving importer, filter and view operations} \label{fig:M1}
\end{figure}

\section{InterActive Computing service (IAC)}
% FABIO, probably let's move it above somewhere closer to introduction
The increasing complexity and interactivity of modern scientific research workflows require compute resources that extend beyond traditional batch processing. In response to this demand, the Fenix\footnote{Fenix, \url{https://fenix-ri.eu/}} infrastructure introduces InterActive Compute services (IAC) as a core component of its federated service portfolio, targeting the specific needs of the neuroscience community within the Human Brain Project\footnote{Human Brain Project, \url{https://humanbrainproject.eu/}} (HBP) and its successor, EBRAINS\footnote{EBRAINS, \url{https://ebrains.eu/}}. The IAC implementation in Fenix was managed by Cineca and E4 Company on the basis of the ICE4HPC suite\footnote{ICE4HPC, https://www.e4company.com/en/ice4hpc/} by E4, and it is currently up and running on the Galileo 100 cluster at Cineca.

The main benefits for an interactive approach to HPC resources are mainly two.

\begin{enumerate}
    \item Users can access directly to the compute node of an HPC system using a web browser interface (available at  \url{https://jupyter.g100.cineca.it}). This is a quite far away approach with respect to the traditional user experience on HPC clusters, which traditionally involves \textit{ssh} access to a single shared login node, and a batch system allowing job submissions in a queue system. The traditional approach, relying only on command line, inhibits any graphical visualization of results, which is on the contrary very smooth on the interactive web interface.
    \item HPC resources are guaranteed in a near-instantaneous access to allow an immediate interaction via web browser. This represents a different philosophy with respect to the above-mentioned batch approach (where the job start time is unpredictable by the user). This approach adds the possibility of an on-the-fly interaction with the workflow while the system is running \cite{jette2023architecture}, a real-time monitoring of the resource usage and visualization of the intermediate results to make decisions on the following steps of the workflow.
\end{enumerate}

\begin{figure}
\centering
        \includegraphics[width=\columnwidth]{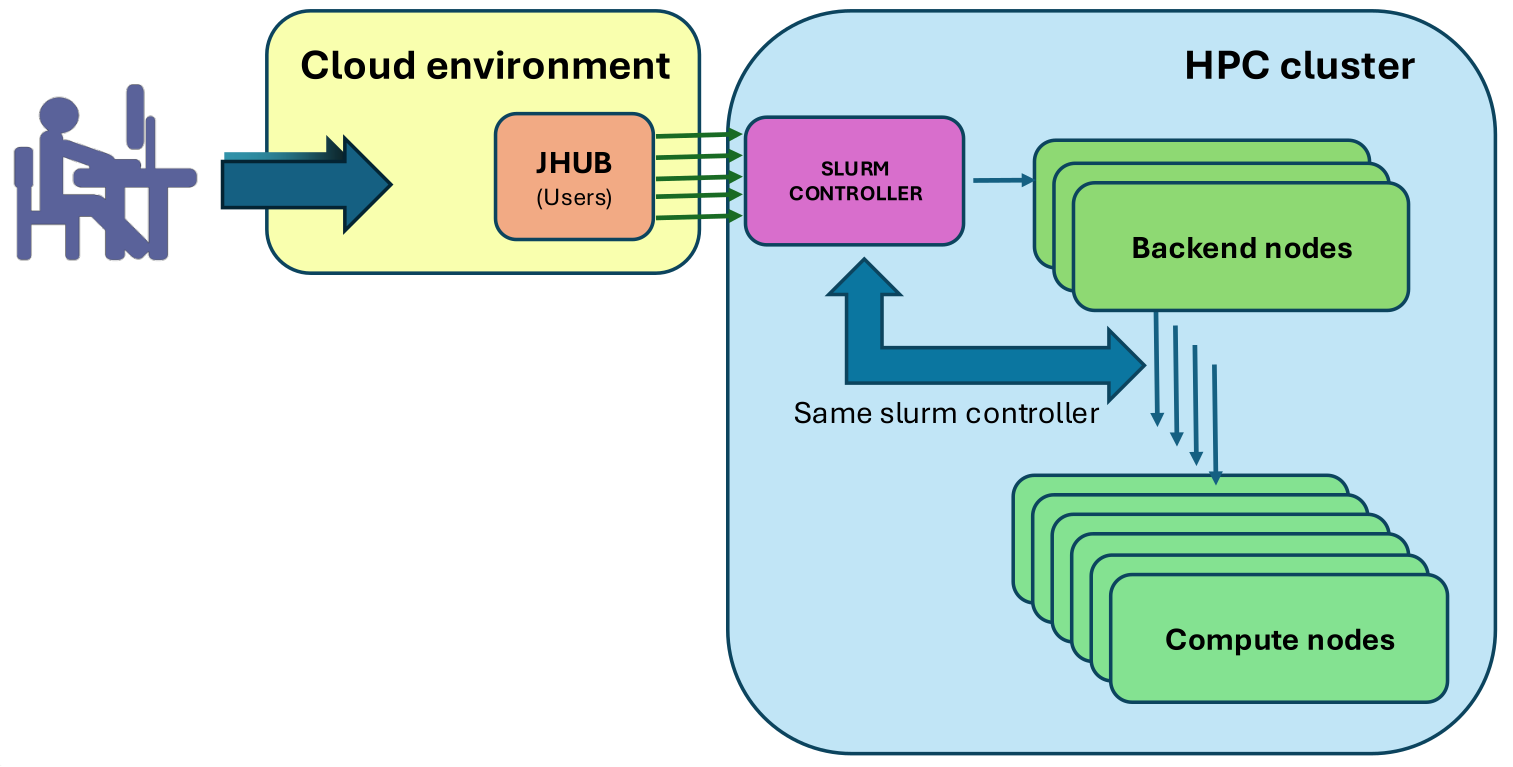}
        \caption{InterActive Computing service implementation at Cineca}
\label{fig:iac_scheme}
\end{figure}

%IAC services provide researchers with ad-hoc access to dedicated compute nodes in a responsive, user-driven manner. Unlike High Performance Computing (HPC) resources managed by batch queuing systems, IAC nodes are designed for real-time engagement, enabling tasks such as interactive data analysis, visualization, and dynamic steering of simulations. This interactive modality is particularly beneficial when working with complex or large-scale datasets, where iterative exploration and visualization are critical to the research process.
The general implementation of the service is depicted in Figure \ref{fig:iac_scheme}. The framework is \changemarker{composed of} three main parts:
\begin{enumerate}
    \item a frontend virtual machine (VM), which exposes the website to the world wide web (it is not exposed from the cluster login nodes for security reasons). The interface asks for user credentials and Two-Factor authentication, then it shows a form to be filled with drop-down menus to request resources to be allocated on the HPC cluster
    \item such requested resources are used to submit a job on the cluster on a dedicated partition, as such a VM is able, exceptionally, to communicate directly with the cluster Slurm controller. The job will run a Jupyter-based \cite{kluyver2016jupyter} server instance (detailed later) which is tunneled to the VM to be exposed externally
    \item the near-istantaneous access is guaranteed since the dedicated Slurm partition (``backend nodes'' in Figure \ref{fig:iac_scheme}) is provided in oversubscription, thus, in the unlucky hypothesis of a fully allocated partition, multiple users would share the same CPU. Oversubscription is not applied to GPUs, which are, on the contrary, allocated exclusively.
\end{enumerate}

%Fenix implemented the IAC service through integration with the Jupyter framework, offering a familiar and extensible environment for researchers. These services are especially suited for workflows requiring human-in-the-loop decision making, where outputs from simulations or data analysis can directly inform subsequent computational steps. % Maybe let's rewrite this stressing that Jupyter is natively familiar with Python, so the added value of the work is to enable visualization and similiar stuff to something (VisIVO) which is not natively Python.
The choice of a Jupyter-based framework for the IAC implementation 
%et- bringed  (??)
brought
several advantages. Firstly, it already includes a server-client approach (Jupyter Single-user) among its possible implementations. Secondly, it includes a plugin system which allowed us to extend the interface to better address the HPC capabilities (e.g. monitoring tools, jupyter-server-proxy plugin to proxy additional http-based servers through Jupyter dashboard). Lastly, it is a well-known environment for many users who approach an HPC environment for the first time, allowing a more user-friendly impact with the infrastructure.

%Importantly, IAC services are not isolated within the Fenix architecture but are deeply interconnected with other compute and data layers. For example, simulations performed on Scalable Compute Services (SCC) can be monitored and influenced through IAC nodes, while intermediate or final data products may be stored in Active or Archival Data Repositories (ACD/ARD) for further examination. This interoperability supports end-to-end workflows within a distributed infrastructure and aligns with Fenix’s goal of offering flexible, federated, and user-friendly services across multiple European research centers.

%In the context of brain research, IAC services enable fine-grained exploration of simulation results, assist in the manual inspection of large neuroimaging datasets, and facilitate collaborative development of analytical tools. Their deployment is a direct response to the evolving computational paradigms in neuroscience, where agility and interactivity are increasingly essential.

\changemarker{The security of the service is strongly improved by the fact that all the operations which might be critical to run on the HPC cluster are delegated to a VM running on Cineca cloud infrastructure. The backend part running on the HPC cluster does not need root privileges, and it works like an ordinary SLURM job in user space. In addition, user login is enforced via 2FA authentication, and constant monitoring of potential security issues, as well as periodic vulnerability assessments and penetration tests (VAPT), are currently performed.}

\section{Methodology}
\label{sec:methodology}

%\subsection{Overview (TBC)}
% ALL
% added by umer sample text

%% Let's stress that:
%% - with IAC you can access directly to compute nodes out of the box
%% - IAC is based on ICE4HPC suite by E4 Analytics, which is deployed via ansible on both the frontend side (VM) and the backend side (compute nodes)

%The interactive computing service offers an alternative way to access computational resources. It is available through a web browser and features an enhanced JupyterLab interface. When users request resources via the browser, they are allocated to a dedicated set of nodes, each equipped with GPUs. These GPUs are assigned exclusively to individual users and are not shared upon request. However, users may share CPUs if the system is fully operational, ensuring quick access without delays. Currently, this service is accessible on Galileo100.

%% In the following let's stress that:
%% - Visivo is not a Python code, so the innovation here is to run any generic "command line based" software via web browser
%% - that this is particularly valuable for a code like Visivo, which is a post-processing tool which produces images; thus this approach enhances the usability, allowing to display images via browser much more easily than the traditional cli approach
%% - that this is an exemplary case in which cloud + hpc interaction strongly improves the user experience
Integrating the VisIVO kernel into the InterActive Computing service simplifies user workflow, eliminating the need for manual module loading. Previously, users logging into the Galileo 100 (G100) cluster had to rely on \textit{ssh} and command line interface to enable VisIVO functionalities. Still, with such an approach the generated PNG files were not easily accessible within the cluster. On the other hand, in the alternative approach we tested in this work, we created a dedicated Jupyter kernel for VisIVO inside the InterActive Computing interface. The VisIVO kernel is now pre-configured, allowing users to log into the InterActive Computing service and select the VisIVO kernel, which automatically loads the VisIVO module. It enables seamless execution of VisIVO commands without additional configuration, ensuring that all generated PNG files are instantly visible within the web interface. The integration simplifies the process and enhances efficiency, allowing users to focus on their scientific visualization tasks without dealing with command-line complexities.

%% We need to rewrite the followings, but showing that in production it is done via ansible accordingly to the ICE4HPC requirements

%Create a customized VisiVO kernel and display it on the Jupyter launcher. Then, open a terminal from the launcher on the Jupyter dashboard. 
%The process begins by initializing Conda to ensure it is correctly loaded into the shell. It configures your shell to properly initialize Conda when you start a new terminal session. A new environment named visivo is created with ipykernel. Activate the visivo environment. visivo Conda environment as a Jupyter kernel, making it available in Jupyter Notebook or JupyterLab with the display name VisiVO. A new environment will now appear in the dashboard launcher.

%To customize the VisiVO kernel, a new JSON configuration file defines how JupyterLab executes the kernel when selected from the launcher. This file specifies the commands that Jupyter runs when the VisiVO kernel is initiated. A wrapper script (wrapper.sh) is also created to load the VisiVO module before launching the kernel. This script ensures that all required dependencies are correctly initialized. Once wrapper.sh is created, it is made executable so Jupyter can invoke it seamlessly. Finally, the kernel configuration file is modified to include the wrapper command, ensuring that the VisiVO module is automatically loaded when users select the VisiVO kernel from the Jupyter dashboard. This setup provides a streamlined experience, eliminating the need for manual module loading while ensuring that all required components are properly initialized.

\subsection{VisIVO wrappers}
%UMER

% VisIVO is a compiled code (C/C++) thus it couldn't be run on Jupyter out of the box
%VisIVO, a compiled code written in C/C++, is at odds with Jupyter, a platform that predominantly executes Python. VisIVO's C/C++ code requires conversion into machine code before execution, while Jupyter effortlessly supports interpreted languages like Python, enabling interactive code execution without the need for compilation. This sharp contrast highlights the incompatibility challenge, necessitating a communication method for Python to interact with VisIVO's compiled functions.

%VisIVO is written in C/C++, thus its compilation provides binary executable. Jupyter doesn't provide a way to integrate binaries executables in its interface, whereas it provides out-of-the-box support for adding interaction with custom Python environments. Thus, we chose to implement a Python wrapper to connect Python and the compiled VisIVO binaries. To execute VisIVO commands within a Python environment,
% Grammarly Improved Version Below
VisIVO is implemented in C/C++; thus, its compilation provides binary executables. Jupyter does not provide a way to integrate binary executables in its interface, whereas it provides out-of-the-box support for adding interaction with custom Python environments.
Thus, we implemented a Python wrapper\footnote{VisIVO Python Wrapper: \url{https://github.com/VisIVOLab/VisIVOPythonWrapper}} to connect Python and the compiled VisIVO binaries to execute VisIVO commands within a Python environment.
% Thus we needed to write some Python wrapper to run Visivo Commands. We used subprocess module from Python standard library
%A Python wrapper is required

%we loaded necessary modules from the Python standard library, such as subprocess. The subprocess module, a crucial part of handling system processes, allows outside commands to run, including those needed for VisIVO. It enables you to directly interact with system processes, giving you control over executing commands, managing input and output, and monitoring the process activity. This integration ensures that VisIVO tasks can be executed at high speed within the Python wrapper. As a result, it enhances the efficiency of the integration and the speed of running VisIVO tasks.

%We rely mostly on the standard library from Python; in particular, we use the subprocess package to run the underlying VisIVO commands, with the idea of keeping track of all the details of the executions via the logging package. This is also done in view of the future developments we're planning (see section \ref{sec:conclusions}).

% Grammarly Improved Version Below
We rely mainly on the standard library from Python; in particular, in this work we used the \textit{subprocess} package to run the underlying VisIVO commands. The idea is to keep track of all the details of the executions via the \textit{logging} package, which is also important in view of the future developments \cite{grinberg2018flask} we are planning (see Section \ref{sec:conclusions}).

% Everything is splitted in different functions to reproduce the three main commands of Visivo (Visivo importer/filter/viewer). This is done with the idea of using decorators in the future for those functions and enhance Visivo functionalities. We can describe here how the code is implemented.
The Python wrapper program organizes functions to manage the three main VisIVO commands: VisIVOImporter, VisIVOFilter, and VisIVOViewer. 
%These commands represent the core functionalities of the code. % VisIVOImporter is responsible for data importation, VisIVOFilter handles data filtering, and VisIVOViewer handles data visualization.
%This structure ensures that each function focuses on a specific command and will allow us to decorate such functions in the future to tune the specific capabilities of each command to our purpose.
This structure ensures that each function focuses on a specific command
%and will allow us to decorate such functions in the future to tune each command's specific capabilities to our purpose. 
%%%% VisIVOImporter, VisIVOFilter, and VisIVOViewer collect input arguments and pass them to the shared _corefunction, where the central processing logic for each individual is handled. This code enables reusability, centralizing the core functionality while keeping the interface functions focused.
%The core function acts as the backbone of the program, dynamically creating command-line arguments by processing various flags and options. When necessary, it supports both standard execution and MPI-based parallel execution. The functions importer, filter, and viewer serve as entry points, each invoking the core function with the appropriate command. This program simplifies future functionality extensions, making adding new features easy, and ensures the code's adaptability through decorators.
%Nevertheless, each of these function embeds a core function which covers the common capabilities of running cli commands, keep track of the logging and match the arguments of the functions with the allowed options of the underlying command.
% Grammarly Improved Version Below
%Nevertheless, each of these functions 
%and embeds a core function that 
covering the common capabilities of running CLI commands, keeping track of the logging, and matching the functions' arguments with the allowed options of the underlying command.
%% insert snippet here
The above three functions are organized as follows (just \texttt{importer} is shown here below as an example): the options from the VisIVOImporter binary are converted as arguments of the function:
\begin{pythonscript}
 def importer(input_file, *flags, mpi=False, tasks=None, fformat=None, out=None, volume=None, compx=None, [...]):
\end{pythonscript}
and after that each of them has been processed to check the variable type, via an \texttt{options} dictionary:
\begin{pythonscript}
        options = {
        "fformat": (fformat, str),
        "out": (out, str),
        "volume": (volume, str),
        "compx": (compx, float),
        "compy": (compy, float),
        "compz": (compz, float),
        "sizex": (sizex, float),
        [...]
\end{pythonscript}
Such dictionary is passed to a function which is in charge to pipe such options to the VisIVO commands:
\begin{pythonscript}
 command = "VisIVOImporter"
 _corefunction(command, input_file, *flags, mpi=mpi, tasks=tasks, options=options)
\end{pythonscript}
We preferred such approach to other more compact possibilities because we choose to hide as much as possible the VisIVO CLI in the background to the user, and thus we preferred to list all the possible options in the function arguments, instead of embedding them in a single optional list of tuples. %%% N.B. Make it simpler!

The main steps of the \texttt{\_corefunction} are the following: first it checks the type of the options values passed from the outer function:
% Let's probably omit the following
\begin{pythonscript}
 worked_options = {}
 if options:
     for key, (val, expected_type) in options.items():
         if val is not None:
             if not isinstance(val, expected_type):
                 raise TypeError(f"'{key}' must be of type {expected_type.__name__}")
             if expected_type is bool:
                 if val:  # add only if True
                     worked_options[key] = val
             else:
                 worked_options[key] = val
\end{pythonscript}
Then, it converts the boolean variables into options without values, and the other ones into options with values:
\begin{pythonscript}
 for option, value in worked_options.items():
     if isinstance(value, (list, tuple)):
         command.append(f"--{option}")
         command.extend(map(str, value))
     elif value and type(value) != bool:
         command.append(f"--{option}")
         command.append(str(value))
     else:
         if value != False:
             command.append(f"--{option}")
\end{pythonscript}
Finally, the VisIVO command to be run is completed adding the input file:
\begin{pythonscript}
    command.append(input_file)
\end{pythonscript}
Such a command is run either via MPI or serially, depending if the \changemarker{variables \texttt{mpi} and \texttt{tasks}} are set or not:
\begin{pythonscript}
 if mpi:
     mpi_command = ["mpirun", "-n", str(tasks)] + command
     logging.debug(f"Running MPI command: {mpi_command}")
     result = subprocess.run(mpi_command, capture_output=True, text=True)
 else:
     logging.debug(f"Running command: {command}")
     result = subprocess.run(command, capture_output=True, text=True)
\end{pythonscript}
Other minor portions of codes are here skipped for the sake of brevity.
% Wrapping Visivo commands via python also allowed to keep detailed log of the actions that the wrapper performed (link with the future section about future developments with flask, since this is useful in a server-like fashion) 
% LET'S MOVE THE FOLLOWING WHEN TELLING ABOUT FLASK
%Wrapping VisIVO commands in Python simplifies execution and enables comprehensive logging of all actions performed by the wrapper. Using Python’s built-in logging module, the program systematically records each command execution, parameters, and results. This detailed logging is crucial for debugging, performance monitoring, and workflow transparency. This structured logging approach is also exciting for future developments, particularly in integrating VisIVO with web-based frameworks like Flask. In a server-like environment, such as one powered by Flask, having detailed logs allows administrators to track user interactions and optimize resource usage efficiently.
% We added also the Pillow module to display png images inside the basic Visivo kernel, the reason is to view the image in command line is difficult and take a lot of time to visualize it. 

VisIVO kernel integrates the \textit{Pillow} module to improve image visualization efficiency, especially for PNG files. Viewing images directly from the command line can be time-consuming, often requiring external tools or manual file opening. By incorporating \textit{Pillow}, users can display images seamlessly within the VisIVO Python environment, eliminating the need for additional steps. This enhancement significantly speeds up the workflow and provides the convenience of immediate image rendering within Jupyter notebooks, making analyzing graphical output easier.

% In order to run visivo via subprocess the visivo binaries has to be availiable in the PATH variable of the bash shell which executes the Visivo python kernel (see deployment section)
%To run VisIVO using the subprocess module in Python, it's crucial that the VisIVO binaries are available in the PATH variable of the bash shell that executes the VisIVO Python kernel.% The PATH variable, an environment variable that guides the system in locating executable files, is of utmost importance in the execution process. If the VisIVO binaries are not in this variable, the system will not recognize the VisIVO commands, leading to execution errors. To ensure proper execution, users must add the directory containing the VisIVO binaries to the PATH before running the Python script. For a more detailed guide on configuring the environment, please refer to the deployment section of the documentation.

\subsection{Deployment}
% FABIO
% deployment is achieved via ansible consistently with what it's done for ICE4HPC by E4 Analytics
% WE NEED TO MENTION SOMEWHERE SOME DETAILS ABOUT THE SPACK INSTALLATION OF VISIVO
% it relies on mamba environment creation and add a custom json file for the kernel management
The environment creation involves four steps:
\begin{enumerate}
    \item Spack module compilation for VisIVO
    \item \textit{conda} environment creation via \textit{Ansible} on the IAC backend nodes
    \item installation of the Python wrappers from \changemarker{the GitHub repository}
    \item customization of \textit{ipykernel}.
\end{enumerate}
The first step involves a system-wide installation of VisIVO, compiling it via Spack \cite{gamblin2015spack}. A Spack module was created forcing OSMesa interface in all the graphical dependencies (notably VTK and Glew). This is a crucial step to let the IAC service  correctly generate the images to be displayed, since there is no windowing system active on the IAC service (only the web interface is up and running) and thus an off-screen rendering is needed.

After this step, a \textit{conda} environment is created via \textit{Ansible} \cite{hochstein2017ansible}. Inside this, the Python wrappers described in Section \ref{sec:methodology} run. There is no particular need of other dependencies since the Python wrappers just rely on the Python standard library, but other Python packages were added to ease the user experience on displaying images and plots (e.g. Pillow, Numpy-based libraries, Matplotlib). This is where Python adds a clear advantage to the traditional VisIVO experience, since from the same environment the users can perform their data analysis relying on both the VisIVO tools and the traditional Python tools, as well as displaying the generated images on the fly via e.g. \textit{IPython} display functionalities inside the Jupyter web interface.

\textit{Ansible} playbooks deploying the VisIVO virtual environment install in parallel the software on the local storage of each one of the computational nodes involved in the Slurm partition dedicated to the IAC service. This is preferred over an installation relying on the parallel file system of the cluster since several tests we performed with the latter approach showed poor performances in the login phase of the service; this is luckily due to the poor management of large numbers of small-sized files by the parallel file system, which is the case of the multiple backend conda environments that has to be read by the Jupyter server initialization procedure. The deployment then relies on an Ansible inventory \changemarker{which explicitly lists 
%et: involves (??)
%et- in its inventory (??)
all the backend nodes participating in the service, and it is executed from the login node of the system (which acts as Ansible controller node).}

After this step, the ipykernel interface which initializes the VisIVO kernel inside the Jupyter interface is modified  to run a bash prolog, which allows to load the spack module containing the VisIVO installation and to set the environment variables it needs.

\section{\changemarker{Use Case Applications}}

\subsection{Workflows description}
\label{sec:workflows}

In this section we demonstrate the integration of the VisIVO Server with the IAC through the three sample workflows described in \cite{sciacca_2024_13858498} and briefly reported hereafter.

\subsubsection{Workflow 1}

The first workflow imports a sample TXT file containing cosmological particles\changemarker{'} positions employing the following command:

\begin{bashscript}
    VisIVOImporter --fformat ascii clusterfields4.ascii
\end{bashscript}

and visualizes it using a data points rendering:

\begin{bashscript}
VisIVOViewer -x X -y Y -z Z --scale --glyphs pixel VisIVOServerBinary.bin
\end{bashscript}

\subsubsection{Workflow 2}
\label{wf2}

The second workflow imports a sample snapshot of a cosmological simulation in GADGET format containing cosmological particles' positions of the GAS, HALO and STARS. The importing of the simulation is executed using MPI/OpenMP because the VisIVO GADGET importer has been recently implemented to be run of HPC infrastructures scaling on multi-nodes and multi-cores when available. This is a sample command when running on 2 cores and 4 nodes:

\begin{bashscript}
export OMP_NUM_THREADS=2
mpirun --np 4 VisIVOImporter --fformat gadget --out NewTable --file snapdir/snap_091.0
\end{bashscript}

The HALO particles in VBT format are then filtered to compute the particle densities using the \textit{pointproperty} filter on a 64X64X64 resolution. 

\begin{bashscript}
VisIVOFilter --op pointproperty --resolution 64 64 64 --points POS_X POS_Y POS_Z --append --outcol density --file NewTableHALO.bin
\end{bashscript}

Finally, we visualize the HALO particles using a data point rendering and the density field for the color palette.

\begin{bashscript}
VisIVOViewer --x POS_X --y POS_Y --z POS_Z --color --colorscalar density --colortable volren_glow --logscale --out VisIVOServerImage NewTableHALO.bin
\end{bashscript}

\subsubsection{Workflow 3}

The third workflow employs the same sample snapshot of the cosmological simulation imported in the workflow reported in Section \ref{wf2} but, differently from Workflow 2, it creates a volume from the particles position using the \textit{pointdistribute} filter which produces a density field distributed and divided for the volume voxels.

\begin{bashscript}
VisIVOFilter --op pointdistribute --resolution 64 64 64 --points POS_X POS_Y POS_Z --out densityvolume.bin --file NewTableHALO.bin
\end{bashscript}

The final volume is then visualized using a volume rendering algorithm.

\begin{bashscript}
VisIVOViewer --volume --vrendering --vrenderingfield Constant --color --colortable volren_glow --showlut --out img --file densityvolume.bin 
\end{bashscript}

%\subsection{Computing Infrastructure}

\subsection{Interactive Notebooks and Rendering results}

% \begin{figure}
%     \centering
%     \includegraphics[width=\linewidth]{iac_laucher.png}
%     \caption{IAC Launcher}
%     \label{fig:iac-launcher}
% \end{figure}

%The VisIVO module is available to be launched as a  Notebook or Console environment, see Figure \ref{fig:iac-launcher}.

\begin{figure}
    \centering
    \includegraphics[width=\linewidth]{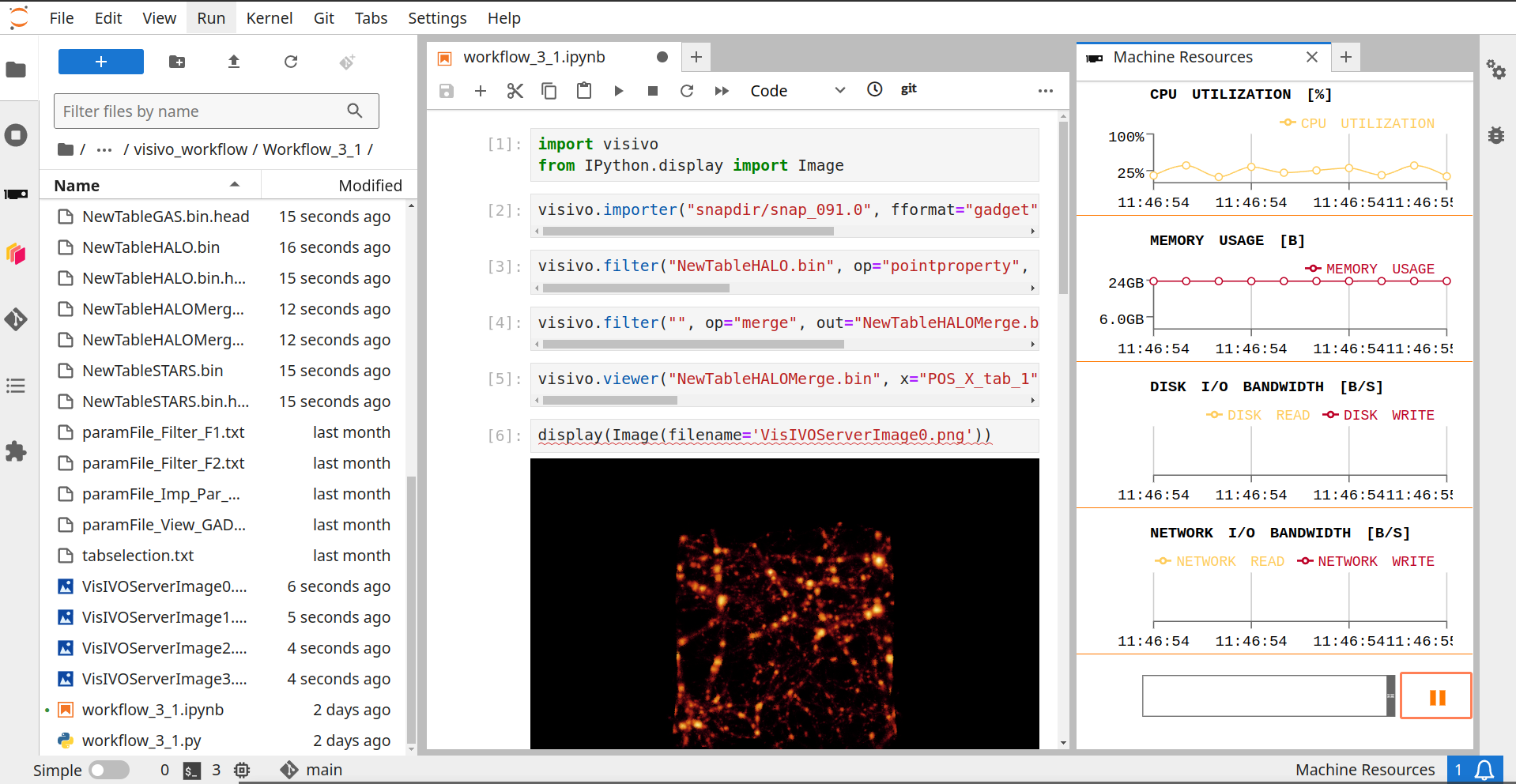}
    \caption{Jupyter notebook running VisIVO via web browser on a compute node of Galileo 100 cluster}
    \label{fig:iac-visivo}
\end{figure}

The VisIVO module is available to be launched as a  Notebook or Console environment, see Figure \ref{fig:iac-visivo}.

The interactive notebooks require the VisIVO wrappers module to be loaded.

\begin{pythonscript}
    import visivo
\end{pythonscript}

\subsubsection{Workflow 1 notebook}

The first workflow is implemented using the following commands:

\begin{pythonscript}
    visivo.importer("clusterfields4.ascii", fformat="ascii")
    visivo.viewer("VisIVOServerBinary.bin", x="X", y="Y", z="Z", scale=True, glyphs="pixel")
\end{pythonscript}

\subsubsection{Workflow 2 notebook}

We implemented the notebook for the second workflow by using the following commands:
\begin{pythonscript}
    visivo.importer("snapdir/snap_091.0", fformat="gadget", out="NewTable", mpi=True, tasks=4)
    visivo.filter("NewTableHALO.bin", op="pointproperty", resolution="64 64 64", points="POS_X POS_Y POS_Z", append=True, outcol="density")
    visivo.viewer("NewTableHALO.bin", x="POS_X", y="POS_Y", z="POS_Z", color=True, colorscalar="density", colortable="volren_glow", logscale=True, out="VisIVOServerImage")
\end{pythonscript}

In this second notebook we can test the capability to execute the importer command using MPI with 4 processes.

\subsubsection{Workflow 3 notebook}
\begin{figure}[h]
\centering        \includegraphics[width=\columnwidth]{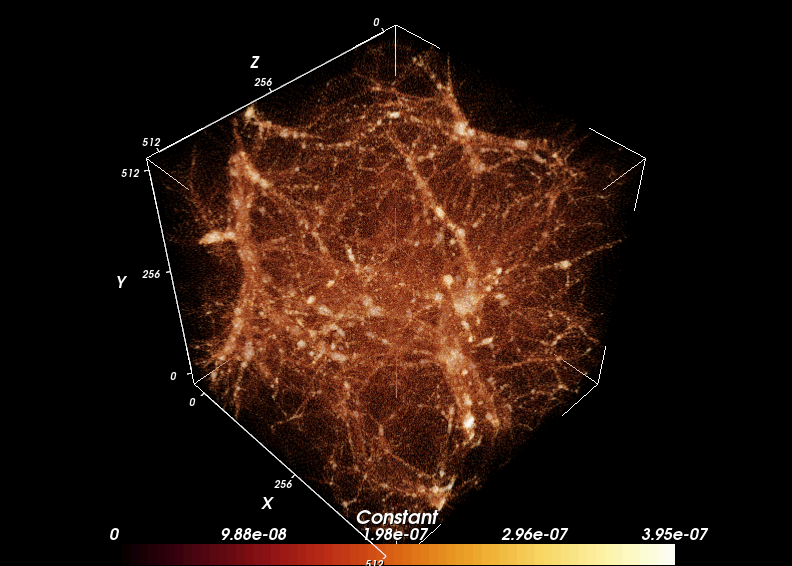}
        \caption{VisIVO volume rendering of HALO particles of GADGET simulation test output}
\label{fig:visivo_render}
\end{figure}

The notebook for the third workflow exeutes the following commands:
\begin{pythonscript}
    visivo.importer("snapdir/snap_091.0", fformat="gadget", out="NewTable", mpi=True, tasks=4)
    visivo.filter("NewTableHALO.bin", op="pointdistribute", resolution="64 64 64", points="POS_X POS_Y POS_Z", out="densityvolume.bin")
    visivo.viewer("densityvolume.bin", volume=True, vrendering=True, vrenderingfield="Constant", color=True, colortable="volren_glow", showlut=True, out="img")
\end{pythonscript}
Figure \ref{fig:visivo_render} shows the rendering results of the GADGET snapshots used in workflows 2 and 3. It presents a volume rendering of the density of GADGET's HALO particles.

\section{Conclusions and Future Works}
\label{sec:conclusions}

This paper presented the innovative incorporation of VisIVO into the InterActive Computing service to allow interactive analyses and visualizations abstracting the underlying large-scale HPC resources.

This activity comprised the implementation of Python wrappers that will be further used in future works to expand the development toward different directions. The integration with Jupyter can be extended to enrich the functionalities or by adding customized functions which were not present in the native command line VisIVO implementation, to exploit the interactive capabilities at most. For instance, specific functions to display images with embedded plots might be easily added relying on the wide availability of Python graphical tools.  Another possible development which still exploits the functions-based structure and avoids once again the need of a VisIVO local installation might be to utilize a REST API built via Flask library to enable the remote execution of VisIVO on a server
%. The main goal would be 
to expose the application's functionality through HTTP endpoints, and provide an interactive interface that abstracts the complexity of backend operations. 
%This is easily achieved via Flask decorators to be applied on the already existing functions, without the need to re-factor the code. %Authentication might be easily handled in Python for instance using JWT tokens, and the logging capabilities of the code might be extended to keep logs on the server in order to track the remote activity by the users.

%The program implements detailed logging and a decorator-friendly structure, allowing better tracking of command executions, parameters, and results. The logging system will help monitor API requests, diagnose errors, and improve debugging, while the decorator-based design will enhance code maintainability. These features provide immediate benefits and pave the way for the future integration of VisIVO into a web-based system, promising greater accessibility and efficiency for users working with large datasets or collaborative environments.

\section*{Acknowledgment}
The work is supported by the EuroHPC JU under grant agreement No 101093441 and by the Spoke 1 ``FutureHPC \& BigData'' of the ICSC – Centro Nazionale di Ricerca in High Performance Computing, Big Data and Quantum Computing, funded by NextGenerationEU.



%\bibliographystyle{IEEEtran}
\bibliographystyle{IEEEtran}
\bibliography{main}

\end{document}